\documentclass[twocolumn,prl,showpacs,superscriptaddress]{revtex4}

\usepackage{graphicx}
\usepackage{dcolumn}
\usepackage{bm}
\usepackage{amssymb}
\usepackage{amsmath}

\begin{document}

\title{Modelling the atomic structure of very high-density amorphous ice}

\author{J. K. Christie}
\email{jkc25@cam.ac.uk}
\affiliation{Department of Chemistry, Cambridge University, Lensfield Road, Cambridge, CB2 1EW, UK}
\author{M. Guthrie}
\altaffiliation[Current affiliation: ]{Centre for Science at Extreme Conditions, Edinburgh University, Edinburgh, EH9 3JZ, UK}
\affiliation{Spallation Neutron Source, Oak Ridge National Laboratory, Oak Ridge, TN 37831}
\affiliation{Argonne National Laboratory, Argonne, IL 60439}
\author{C. A. Tulk}
\affiliation{Spallation Neutron Source, Oak Ridge National Laboratory, Oak Ridge, TN 37831}
\author{C. J. Benmore}
\affiliation{Argonne National Laboratory, Argonne, IL 60439}
\author{D. D. Klug}
\affiliation{National Research Council of Canada, Ottawa, Ontario, Canada KA1 0R6}
\author{S. N. Taraskin}
\affiliation{Department of Chemistry, Cambridge University, Lensfield Road, Cambridge, CB2 1EW, UK}
\author{S. R. Elliott}
\affiliation{Department of Chemistry, Cambridge University, Lensfield Road, Cambridge, CB2 1EW, UK}

\date{\today}

\begin{abstract}
The structure of very high-density amorphous (VHDA) ice has been modelled 
by positionally disordering three crystalline phases, namely ice IV, VI 
and XII. These phases were chosen because only they are stable or 
metastable in the region of the ice phase diagram where VHDA ice is 
formed, and their densities are comparable to that of VHDA ice. An 
excellent fit to the medium range of the experimentally observed 
pair-correlation function $g(r)$ of VHDA ice was obtained by introducing 
disorder into the positions of the H$_{\textrm{2}}$O molecules, as well as 
small amounts of molecular rotational disorder, disorder in the O--H bond 
lengths and disorder in the H--O--H bond angles. The low-\emph{k} behaviour 
of the experimental structure factor, $S(k)$, is also very well 
reproduced by this disordered-crystal model. The fraction of each phase 
present in the best-fit disordered model is very close to that observed in 
the probable crystallization products of VHDA ice.  In particular, only
negligible amounts of ice IV are predicted, in accordance with experimental
observation.
\end{abstract}


\maketitle


The study of water has attracted much attention  
\cite{water franks,petrenko book}.
This is not only because of its importance in nature, but also because
it has a number of unusual properties. In particular, it has a very
rich phase diagram. In addition to liquid water, thirteen
crystalline phases of ice are known to exist, as well as at least three
amorphous, or glassy, phases \cite{petrenko book}. Mishima \emph{et
al}.~\cite{mishima} were the first to synthesize high-density
amorphous (HDA) ice, and the relationships between this phase, low-density
amorphous (LDA) ice, and liquid water have been well studied 
\cite{possibly debenedetti}.
Recently, another distinct phase of amorphous ice has been 
made, by heating HDA ice isobarically at $1-2$ GPa from
77K to $\sim$ 165K. The new phase of ice can be recovered at 77K and 
1 bar \cite{loerting et al}. This
new phase is even higher in density than HDA ice, and has become known as 
very high-density
amorphous (VHDA) ice \cite{loerting et al}. VHDA ice, and its relationship
to HDA and LDA ice, is of considerable interest in studies of polyamorphism
\cite{angell review} (in which a substance exhibits more than one 
amorphous phase), and its structure is also relevant to the structure 
of liquid water.

Water molecules form strong hydrogen bonds to neighbouring water molecules,
resulting in a roughly tetrahedral molecular arrangement. 
The hydrogen-bond-induced flexibility
of this structure is the main reason for the many different phases
of ice. HDA and VHDA ice exhibit closer
packing of the water molecules.  Each water molecule is still bonded to
four nearest neighbours but, between 
the first and second neighbour shells, there is
approximately one extra molecule in HDA ice, and two extra molecules
in VHDA ice.  This increase in coordination number is observed
experimentally \cite{soper,finney 1,finney 2} and attributed to occupation
of the {}``lynch pin'' location by interstitial water molecules which are 
not hydrogen bonded to the origin molecule. Recent MD simulations support this
interpretation \cite{giovambattista}, and suggest that VHDA ice is
more stable than HDA ice, and that it is the amorphous analogue of the
postulated high-density liquid water structure \cite{mishima review}.

The medium-range structure of amorphous materials, that
is how atomic positions are correlated over several interatomic
distances, is of much interest \cite {SRE nature}.  
VHDA ice is exceptional among amorphous materials in exhibiting
very pronounced medium-range order (MRO).  The orientationally averaged
real-space pair-correlation function, $g(r)$, for
$r\gtrsim5$\AA, consists of damped real-space
oscillations extending to $\sim20$\AA, with only one significant period, $D$ 
\cite{guthrie et al}.  The orientationally averaged structure
factor, $S(k)$, measured by neutron and X-ray diffraction \cite {guthrie
et al}, exhibits an intense, sharp peak at $k_{0}=2.29(2)$\AA$^{-1}$ in the
neutron data, and $2.30(2)$\AA$^{-1}$ in the X-ray data, this peak being the
dominant feature (particularly for the neutron data).  Since $g(r)$ and $S(k)$
are related by a Fourier transform, the period of the oscillation in $g(r)$
is related to the position of the peak in $S(k)$ by $D=2\pi/k_{0}$.

The structure of amorphous materials has recently been modelled by 
adding Gaussian positional disorder
into the atomic structure of crystalline phases of these
materials \cite{JKC 1,JKC 2}. These disordered crystals
gave good representations
of the MRO of some amorphous materials, but not 
necessarily the short-range order (SRO). In
particular, the value of the wavevector transfer of the
so-called first sharp diffraction peak (FSDP, a characteristic peak in the
$S(k)$ of many amorphous materials at low wavevector, 
$k\sim1-2$\AA$^{-1}$) was successfully predicted
in amorphous silicon and vitreous silica. The existence of the FSDP
is often regarded as a signature of MRO \cite{SRE nature}.

For a model crystalline lattice, $S(k)$ consists of a series of 
$\delta$-functions caused by scattering from the various Bragg planes.
The effect of Gaussian positional disorder is to multiply these peaks
by the Debye-Waller factor, $e^{-k^{2}\sigma^{2}}$, where $\sigma$
is the half width of the Gaussian distribution, and
to produce a low-amplitude diffuse scattering \cite{willis and pryor}.
It is clear that the high-\emph{k} peaks will be destroyed first by
introducing disorder, and that for large enough disorder (high
enough $\sigma$), only one peak (that at the lowest value of \emph{k})
will contribute substantially to $S(k)$.

Similarly, for a model crystalline lattice, $g(r)$ consists of 
$\delta$-functions at the various interatomic
distances. The effect of Gaussian positional disorder is to convolve these
peaks with the Fourier transform of the Debye-Waller factor, which
is also a Gaussian \cite{willis and pryor}. For large enough disorder 
(high enough $\sigma$), these Gaussian peaks broaden and combine to 
form damped oscillations
in $g(r)$ with a \emph{single} period \cite{JKC 1, JKC 2}. This functional 
behaviour, of course, results from the Fourier transformation of $S(k)$
having only one significant peak.

The experimentally measured $g(r)$ for VHDA ice 
\cite{guthrie et al}, showing pronounced MRO, characterized by nearly 
periodic oscillations 
extending to large \emph{r},
accompanied by a dominant sharp low-\emph{k} peak in $S(k)$, are
very reminiscent of the simulated data for positionally disordered
lattices \cite{JKC 1,JKC 2}. It was natural to ask whether
the MRO in VHDA ice could be modelled by disordering
crystalline ice.  The structure of
an amorphous phase is most likely to resemble the crystalline phase with a 
similar density that is 
stable at similar temperatures and pressures.  VHDA ice is formed at $\sim$
165K and 1--2 GPa, with density $\rho=1.25$ g/cm$^{3}$ (all densities
are for H$_{\textrm{2}}$O) 
when recovered to 1 bar at 77K \cite{loerting et al}.  
According to the phase diagram, ice VI is 
the only crystalline phase stable in this region \cite{petrenko book}.  Four
other crystalline phases (I$_c$, IV, IX and XII) are metastable 
\cite{petrenko book} and hence their occurrence around these temperatures
and pressures cannot in principle be excluded.
However, ice I$_c$ and IX can be eliminated as 
candidates because their densities (respectively 0.93 g/cm$^{3}$ at 77K 
and 1 bar \cite{petrenko book}, 
and 1.17 g/cm$^{3}$ at 110K and 1 bar \cite{londono et al}) 
are so much lower than that of
VHDA ice, and it is very unlikely that a positionally disordered phase could
have a larger density than the corresponding crystalline phase.  If there were
significant density variations in VHDA ice, then small-angle
scattering would be experimentally observed, which it is not.
It was therefore decided to disorder the structures of 
ice IV ($\rho=1.29$ g/cm$^{3}$ at 260K and 0.5GPa \cite{lobban et al}), VI 
($\rho=1.37$ g/cm$^{3}$ at 225K and 1.1GPa \cite{kuhs et al}) and XII
($\rho=1.29$ g/cm$^{3}$ at 260K and 0.5GPa \cite{lobban et al}), 
in order to try to simulate VHDA ice.
\begin{figure}[t]
\begin{center}\includegraphics[%
  width=2.15in,
  angle=270]{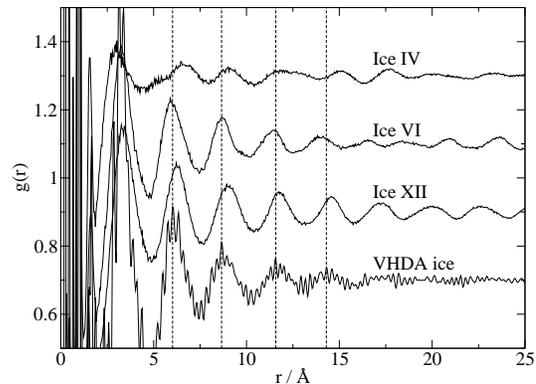}\end{center}

\caption{\label{fig1}The computed pair-correlation function, $g(r)$, 
for positionally disordered ice IV, VI
and XII lattices, compared to experimental data \cite {guthrie et al} 
for VHDA ice. The parameters used to generate the simulated curves 
are those of the best-fit three-phase models, as given in Table \ref{Table1}.
The dashed vertical lines show the peak positions in $g(r)$ for VHDA ice.  It
is clear that the peak positions and periods of oscillation are different for
each phase.  The $g(r)$'s have been offset vertically for clarity.}
\end{figure}

Crystal models (with $\sim3\times10^{4}$ atoms) of each phase
were created. The crystalline atomic positions for ice IV \cite{klotz et al}, 
VI \cite{kuhs et al} and XII \cite{klotz et al} 
were taken from neutron-diffraction data.
The hydrogen atoms were positioned according to the 
{}``ice rules'' \cite{pauling ice rules}:
along each O--O bond there is only one hydrogen atom, which is covalently
bonded (and thus closer) to one oxygen atom, and hydrogen bonded to (and
thus further away from) the other oxygen atom. For these phases, this
means that only half of the sites available for hydrogen atoms are
occupied; this fractional occupancy is 
observed experimentally \cite{klotz et al,kuhs et al}.
An algorithm due to Buch \emph{et al}.~\cite{buch et al} was used
to position the hydrogen atoms on their crystalline sites. The models
were rescaled slightly so that their densities matched that of VHDA ice.

Disorder was then introduced into the models as follows. 
Positional disorder was
introduced by displacing the H$_{\textrm{2}}$O molecules rigidly 
in the $x$-, $y$- and
$z$-directions. Then the H$_{\textrm{2}}$O molecules were rotated around 
the $x$-, $y$- and $z$-axes passing
through the oxygen atom positions, such that the oxygen atoms were
kept stationary.
The hydrogen atoms were then moved along the O--H bond direction, 
to disorder the
O--H bond lengths. Finally, the hydrogen atoms were symmetrically moved in the 
H--O--H plane to disorder the H--O--H bond angles.  In each case, the change in
the relevant quantity (position or angle) was randomly chosen from a Gaussian
distribution centred on zero, with half widths as given in Table \ref{Table1}.
\begin{table*}
\begin{tabular}{|c|c|c|c|c|c|c|c|}
\hline
&
Half width&
Range studied&
ice IV&
\multicolumn{2}{c}{ice VI}&
\multicolumn{2}{|c|}{ice XII}\tabularnewline
&
&
&
3 phases&
2 phases&
3 phases&
2 phases&
3 phases\tabularnewline
\hline
Positional disorder&
$\sigma_{pos}$&
$0.05-0.30$\AA&
0.22\AA&
0.16\AA&
0.16\AA&
0.19\AA&
0.18\AA\tabularnewline
\hline
Rotational disorder&
$\sigma_{\theta}$&
$0.01-0.05$ rad&
0.04 rad&
0.04 rad&
0.04 rad&
0.03 rad&
0.01 rad\tabularnewline
\hline
O--H bond length disorder&
$\sigma_{OH}$&
$0.01-0.04$\AA&
0.03\AA&
0.04\AA&
0.04\AA&
0.02\AA&
0.01\AA\tabularnewline
\hline
H--O--H bond angle disorder&
$\sigma_{HOH}$&
$0.01-0.04$ rad&
0.01 rad&
0.01 rad&
0.01 rad&
0.03 rad&
0.02 rad\tabularnewline
\hline
Volume fraction, \emph{f}&
-&
$0-1$&
0.035&
0.544&
0.549&
0.456&
0.416\tabularnewline
\hline
\end{tabular}
\begin{caption}{\label{Table1}Disorder parameters for the
crystalline phases, ice IV, VI and XII. The ranges of 
values of the disorder parameters and the fractions of each phase,
and the values in the best-fit two- and three-phase models, are also given.  
In each case, the change in the position or angle was randomly taken 
from a Gaussian distribution, centred on zero, with half widths as shown.}
\end{caption}
\end{table*}

Calculated $g(r)$'s for these models were fitted to
neutron-diffraction data 
\cite{guthrie et al} of deuterated VHDA ice, the structure of which 
differs negligibly from the structure of the hydrogenated material 
\cite{tomberli et al}. The neutron
data were used, as they were of a higher quality than the X-ray data. 
Since particularly good agreement with the data at low $r$ was not expected, 
the data for $r<7$\AA~were excluded from the fit. 
The partial pair-correlation functions $g_{OO}(r)$, $g_{OD}(r)$
and $g_{DD}(r)$ of our models were combined to give
the full neutron $g(r)$ as 
$g(r)=0.088g_{OO}(r)+$ $0.418g_{OD}(r)+$ $0.494g_{DD}(r)$,
as in Ref.~\cite{guthrie et al}.

The calculated $g(r)$'s
for the positionally disordered models of each separate phase do not
exactly fit the $g(r)$ of VHDA ice
(see Fig.~\ref{fig1}).

Hence, the structure of VHDA ice was assumed to comprise an inhomogeneous
mixture of regions of positionally disordered ice IV, VI and XII,
of indeterminate size, but with total volume fractions of $f_{IV}$, $f_{VI}$
and $f_{XII}=1-(f_{IV}+f_{VI})$, respectively. The $g(r)$'s 
of the three phases were combined to give a total
pair-correlation function: $g(r)=f_{IV}g_{IV}(r)+$ $f_{VI}g_{VI}(r)
+$ $f_{XII}g_{XII}(r)$, where $0<f_{IV},f_{VI},f_{XII}<1$, 
and e.g.~$g_{IV}(r)$ is the full $g(r)$ of the disordered ice IV lattice. 
For each combination of the $g(r)$'s of the three phases,
a least-squares minimization was performed with respect to 
$f_{IV}$ and $f_{VI}$. 

\begin{figure}
\begin{center}\includegraphics[%
  width=2.15in,
  angle=270]{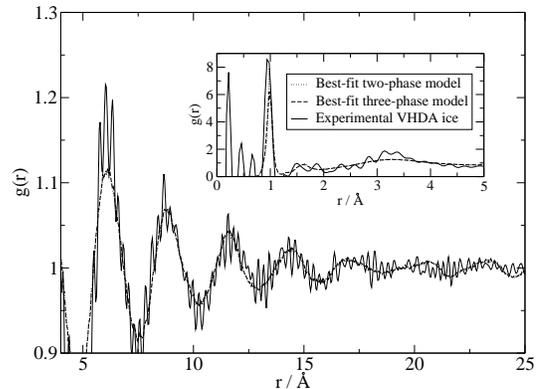}\end{center}

\caption{\label{g(r) figure}The pair-correlation functions $g(r)$ of the
best-fit models consisting of disordered lattices of ice IV, VI and XII 
(dashed line) and ice VI and XII only (dotted line),
compared with the experimentally measured neutron-diffraction $g(r)$
of VHDA ice \cite{guthrie et al} (solid line). 
The disorder parameters of the models are given in Table \ref{Table1}.
The best-fit two- and three-phase models are indistinguishable for
medium-range distances. 
The $g(r)$ for VHDA ice was calculated by Fourier inversion
of the experimentally measured structure factor \cite{guthrie et al}
without smoothing, resulting in considerable noise, and spurious peaks
for $r\lesssim0.8$\AA.}
\end{figure}

The minimization analysis showed that $\sigma_{pos}$
and the volume fractions were the most significant parameters 
in determining the
quality of the fits in the medium range ($r\geq7$\AA).
The other parameters, $\sigma_{\theta}$, $\sigma_{OH}$ and $\sigma_{HOH}$,
were much less important, and only significantly affected $g(r)$ at low
$r$.  The best-fit model (with parameters as given in Table \ref{Table1})
gives a
total $g(r)$ with medium-range oscillations in very good agreement
with those found experimentally for VHDA ice (Fig.~\ref{g(r) figure}),
and very significantly better than for the disordered individual phases
(see Fig.~\ref{fig1}).
The short-range fit is less good; in particular, the heights of the
peaks at $\sim1$\AA~and $\sim3.3$\AA, are not reproduced so well (see inset
to Fig.~\ref{g(r) figure}). 
The height of the latter peak is known to be sample and pressure dependent, so
one would not necessarily expect a good fit.  The best-fit model has a 
least-squares
deviation from the experimental data of 0.0777.  It contains 3.5\% ice IV, 
an amount which is not
resolvable by diffraction studies.  The fitting procedure was repeated for
disordered ice VI and XII only, and the best-fit model 
(Fig.~\ref{g(r) figure}) has a deviation of 0.0779,
which is only 0.25\% more than the deviation for the best three-phase model.
The corresponding $g(r)$'s are indistinguishable (Fig.~\ref{g(r) figure}).
We conclude that the amount
of ice IV needed to model VHDA ice is negligible, and that disordering only 
ice VI and XII produces just as good a description of the MRO of VHDA ice.

\begin{figure}[t]
\begin{center}\includegraphics[%
  width=2.15in,
  angle=270]{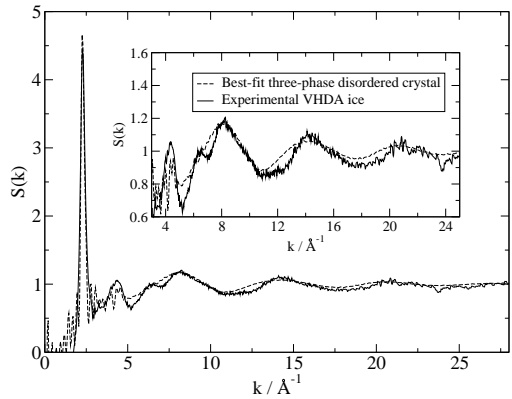}\end{center}

\caption{\label{S(k) pic}The structure factor $S(k)$ of the best-fit model
consisting of disordered lattices of ice IV, VI and XII, compared with
the experimentally measured neutron-diffraction $S(k)$ of VHDA ice 
\cite{guthrie et al}. The parameters of the model are given in 
Table \ref{Table1}.}
\end{figure}

It is also instructive to compare the structure factor $S(k)$ of
the best-fit model 
(calculated by Fourier-inverting $g(r)$) with the experimental
data (Fig.~\ref{S(k) pic}). The height, width and position of
the first peak are fitted very well, which is expected since
this peak governs the medium-range real-space behaviour. The
fit at higher \emph{k} (which corresponds to small $r$) is less 
good, although the peaks are reasonably well reproduced.

When Klotz \emph{et al}. recrystallized high-pressure HDA 
(probably VHDA) ice, the fractions of the phases present were 56\% ice VI
and 44\% ice XII \cite{klotz et al}.  
These are very close to the simulated values of 54\% ice VI 
and 46\% ice XII for the best-fit two-phase model.  This implies that the 
structure of VHDA ice is intimately related to the structures of 
ice VI and XII.  It is also interesting to note that the value of 
$\sigma_{pos}$ in our models (Table \ref{Table1}) is comparable to the thermal 
mean-square  atomic displacement, which for these phases is 
$\sim0.2$\AA~\cite{klotz et al, kuhs et al}.

Attempts were made to improve the fit at short range, by fitting separately
to the low-\emph{r} region of $g(r)$, and to the medium- and high\emph{-k}
region in $S(k)$, whilst varying the short-range disorder parameters
$\sigma_{\theta}$, $\sigma_{OH}$, $\sigma_{HOH}$ and volume fractions,
but none of these attempts improved
the fit of the model. Some of this discrepancy may be attributed
to effects at the boundaries between regions of disordered
ice phases proposed in the VHDA ice structure. Our models
take no account of these boundaries, yet if the ice structure
consists of positionally disordered domains of different crystal phases,
then the interfacial regions will not necessarily resemble any of the
phases. It is likely that any such distortion
will be fairly local, extending over maybe two nearest-neighbour bond
lengths, confining these surface effects to low \emph{r}.

The fact that the disordered crystal model is able to
reproduce the medium-range structure, but not particularly that 
of the first- and
second-peak region in $g(r)$, for VHDA ice may be an important indication
of structural differences between the amorphous and crystalline
phases. It is well known that the 4--8 \AA~range in
amorphous ice is very sensitive to changes in pressure, and X-ray
diffraction studies of VHDA ice have shown
that this region contains two peaks which are most likely related
to the start of the formation of non-bonded interpenetrating lattices
(similar to those present in ice VI and XII).
A distortion of the local tetrahedral unit in VHDA ice enables an
increased packing of interstitial molecules, suggesting that much
of the pressure-induced frustration in the material occurs largely
in the ordering of the molecules just beyond the first shell. This
frustration arises from competition between the optimum local packing
of molecules and the preferred longer-range topology of the network.
Such high-density structures are very different from that characterized
by the well-defined
first two shells found in LDA ice, which closely resembles that of ice
I$_{\textrm{h}}$.

In conclusion, the
medium-range structure of very high-density amorphous (VHDA) ice 
has been modelled
by adding positional disorder to three crystalline phases of ice, namely
ice IV, ice VI and ice XII.  These phases were chosen 
because they are stable or metastable in the 
region of the phase diagram where VHDA ice is formed, and their densities 
are similar to that of VHDA ice.
For medium-range distances ($r\gtrsim7$\AA),
the fit to the experimentally observed neutron-diffraction $g(r)$
is very good. The fractions of each phase present in the
model which gives the best fit to the experimental data are
very close to those found when a sample, which was likely
to have been VHDA ice, was recrystallized \cite{klotz et al}; in particular, 
the amount of ice IV present is negligible (none was observed experimentally). 
We conclude that the structure of VHDA ice at large
enough distances from an arbitrary origin atom can be well modelled
as an inhomogeneous mixture of positionally disordered ice VI and
ice XII.

JKC would like to thank the Engineering and Physical Sciences Research
Council for financial support.

\end{document}